\documentclass[compsoc,conference,a4paper,10pt,times]{IEEEtran}
\usepackage{cite}
\usepackage{amsmath,amssymb,amsfonts}
\usepackage{algorithmic}
\usepackage{graphicx}
\usepackage{textcomp}
\usepackage{bmpsize}
\usepackage{xcolor}
\usepackage{lipsum}
\usepackage[colorlinks=true,urlcolor=black]{hyperref}

\usepackage{graphicx}
\usepackage{caption}
\usepackage{subcaption}
\usepackage{booktabs}
\usepackage{multirow}
\usepackage{url}
\usepackage{stfloats}

\def\BibTeX{{\rm B\kern-.05em{\sc i\kern-.025em b}\kern-.08em
    T\kern-.1667em\lower.7ex\hbox{E}\kern-.125emX}}

\usepackage[colorinlistoftodos]{todonotes}

\begin{document}

\title{CAN We Trust Your Results? \\A Cross-Dataset Study of Automotive IDS Evaluation}

\author{
\IEEEauthorblockN{Beatrix Koltai\textsuperscript{1}, Gergely Ács\textsuperscript{1,2}, András Gazdag\textsuperscript{1}}
\IEEEauthorblockA{\textsuperscript{1}Laboratory of Cryptography and System Security (CrySyS Lab), \\
Budapest University of Technology and Economics, Hungary\\
\textsuperscript{2}HUN-REN-BME Information Systems Research Group, Hungary \\
\{bkoltai, andras.gazdag, acs\}@crysys.hu}
}

\maketitle

\begin{abstract}
The increasing connectivity of modern vehicles has made securing in-vehicle communication networks a critical challenge. Intrusion Detection Systems (IDS) have been widely studied as a defense mechanism for detecting malicious activities on the Controller Area Network (CAN) bus. However, the evaluation of CAN IDS methods remains difficult due to inconsistencies in experimental setups and the lack of standardized benchmarking frameworks. As a result, reported performance often depends on dataset-specific characteristics and may not reflect how detection methods behave in different environments.
This work introduces a benchmarking framework for consistent evaluation of CAN IDSs across multiple datasets. Using the proposed framework, we integrate seven publicly available CAN IDS datasets collected under different experimental conditions and perform cross-dataset evaluation of five conceptually different IDS approaches. Our results highlight how detection performance can vary significantly across datasets, demonstrating the importance of cross-dataset benchmarking for assessing the robustness and generalization capabilities of CAN IDS methods.

\end{abstract}


\section{Introduction}

The growth of connectivity and software functionality in modern vehicles has significantly increased the importance of securing in-vehicle communication systems. Contemporary automobiles contain numerous electronic control units (ECUs) that interact through internal networks to manage vehicle functions. Among these networks, the Controller Area Network (CAN) bus is the dominant communication protocol. Because the CAN protocol was originally designed without built-in security mechanisms, it is vulnerable to various cyberattacks, particularly as modern vehicles become increasingly connected and software-driven. As a result, intrusion detection systems (IDSs) have become widely studied for detecting malicious activities on automotive networks.

In recent years, a large number of CAN-based IDS approaches have been proposed, ranging from rule-based and statistical methods to machine learning and deep learning techniques \cite{rajapaksha_ai_based_2023, al_jarrah_intrusion_2019}. However, inconsistencies in experimental infrastructure, evaluation methodology, and reported performance metrics make it difficult to reliably assess the effectiveness of these approaches. Prior studies have noted that the lack of standardized benchmarking environments complicates both the adoption of state-of-the-art IDS methods and the comparison of results across different works, as well as the interpretation of their real-world performance in new settings \cite{agbaje_framework_2022, sharmin_benchmarking_2024}. 

Another challenge arises from the datasets used for evaluation. Many studies rely on proprietary datasets that are not publicly available, or use datasets collected under different experimental conditions. Consequently, reported performance often depends on the specific dataset and experimental setup, making results difficult to reproduce or compare directly \cite{pollicino_performance_2024}. Reproduction of prior work is further complicated by the limited availability of implementation details or source code. These issues highlight the need for systematic comparison studies based on publicly available benchmark datasets. Most existing benchmark studies focus on testing multiple IDS approaches on a single dataset. 
While such experiments are obviously needed, they do not reveal whether detection methods generalize across different vehicle environments, attack scenarios, or data collection setups (see Table~\ref{tab:contribitions_comparison}).

Studies in the broader domain of machine learning for computer security have highlighted several methodological pitfalls related to dataset construction and evaluation protocols \cite{arp_pitfalls_2024}. In particular, security datasets often contain environment-specific artifacts introduced during data collection, preprocessing, or labeling, which learning algorithms may rely on instead of capturing the underlying malicious behavior. Even simple or unintended baselines can perform surprisingly well when such artifacts are present. As a result, models evaluated using a single dataset may achieve high reported performance while relying on dataset-specific correlations rather than security-relevant features \cite{geirhos_shortcut_2020, arp_pitfalls_2024}.

In practice, IDS performance may degrade when deployed in environments that differ from those used during development, a phenomenon commonly associated with dataset bias and data drift.
Such degradation can arise from distribution shifts between training and deployment data. 
For example, differences in vehicle platforms, ECU configurations, or driving conditions may alter the distribution of CAN messages, leading to covariate shift between training and evaluation data. Similarly, datasets may contain different proportions or implementations of attack types, resulting in label distribution changes.
These factors can significantly influence the reported performance of IDS methods and limit the validity of comparisons based solely on single-dataset evaluations.
Therefore, cross-dataset evaluation is essential to better understand the robustness and generalization capabilities of IDS approaches.

\begin{table}[t]
\caption{Feature comparison of CAN IDS benchmarking studies.}
\label{tab:contribitions_comparison}
\centering
\setlength{\tabcolsep}{3pt}
\renewcommand{\arraystretch}{1.1}

\begin{tabular}{p{3.3cm}ccccc}
\toprule
Feature &
\cite{blevins_time_based_2021} &
\cite{sharmin_comparative_2023} &
\cite{pollicino_performance_2024} &
\cite{agbaje_framework_2022} &
\textbf{This Work} \\
\midrule

Benchmarking framework 
&  &  & \checkmark & \checkmark & \checkmark \\

Multiple IDS categories 
&  & \checkmark &  & \checkmark & \checkmark \\

Multiple attack families 
& \checkmark & \checkmark &  &  & \checkmark \\

Multiple datasets evaluated
&  &  & \checkmark &  & \checkmark \\

Multiple vehicle platforms
&  &  & \checkmark &  & \checkmark \\

Cross-dataset evaluation
&  &  &  &  & \checkmark \\

\bottomrule
\end{tabular}

\vspace{2mm}
\footnotesize $\checkmark$ indicates that the feature is explicitly supported by the study.
\end{table}

During our experiments, we identified a case where inconsistencies in published performance metrics were caused by an incorrect interpretation of evaluation outputs (see Section~\ref{2_Methods}). Detecting the issue required re-implementing the method and recomputing the metrics, highlighting how difficult it can be to verify reported results. This observation directly motivates the need for standardized and reproducible evaluation procedures, where metric computation is clearly defined and consistently applied across methods.

To address these challenges, this paper introduces a benchmarking framework designed for the consistent evaluation of CAN intrusion detection methods across multiple datasets. The framework enables reproducible experimentation by standardizing dataset representation, evaluation procedures, and IDS integration.
By integrating seven commonly used public CAN intrusion detection datasets collected under different experimental conditions, the proposed framework enables systematic cross-dataset benchmarking. This allows us to investigate whether IDS methods generalize across datasets or whether their performance is strongly influenced by dataset-specific characteristics such as vehicle platform, traffic conditions, attack type, or data collection methodology.

The main contributions of this work are summarized as follows:

\begin{itemize}
\item We propose a unified benchmarking framework for evaluating CAN intrusion detection systems in a reproducible and consistent manner.
\item We integrate seven publicly available datasets with different attack scenarios and vehicle environments into a common representation.
\item We provide a standardized evaluation workflow that enables fair comparison between IDS methods with different detection principles.
\item We perform cross-dataset evaluation on five conceptually different methods to study the generalization capabilities of such CAN IDS approaches across multiple datasets and attack families.
\end{itemize}

Table~\ref{tab:contribitions_comparison} compares the features of existing CAN IDS benchmarking studies with this work. Unlike prior work, our framework supports cross-dataset evaluation while simultaneously considering multiple IDS categories, attack families, and datasets.

\section{Background}\label{B}

This section provides an overview of the CAN traffic and the potential attacks targeting it, and outlines the general concept of the Intrusion Detection Systems chosen for comparison in this study.


\subsection{Controller Area Network}\label{BC}

\begin{figure}[ht]
  \centering
  \includegraphics[width=\columnwidth]{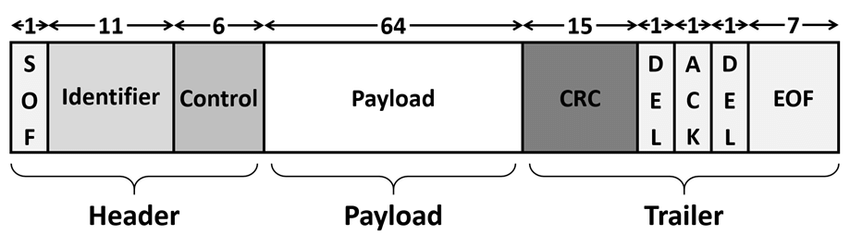}
  \caption{Structure of a CAN frame \cite{kukkala2020indra}.}
  \label{fig:canframe}
 \end{figure}

Modern-day vehicles have a complex internal control system of Electronic Control Units (ECU), each assigned to manage a specific function. The Controller Area Network (CAN) is one of the communication protocols used between these ECU. On the CAN bus, information is transmitted in frames, as shown in Figure \ref{fig:canframe}, containing header with an identifier (ID) and data length code (DLC); payload with the transmitted data, ranging from 0 to 8 bytes; and trailer segments \cite{pollicino_performance_2024, koltai_supporting_2024}.

The payload segment carries various digital and analog signals, and the byte boundaries corresponding to these signals are defined in a DBC file, which is typically proprietary to the manufacturer. As these files are often kept confidential, the signal encoding within the payload is generally unknown without access to the DBC. Consequently, several reverse-engineering techniques have been developed to infer signal structure and encoding from raw CAN traffic, building on methods previously proposed in the literature \cite{Marchetti2019READ,Verma2021CAN-D,Markovitz2017field-classification}. 


\subsection{Attacks on CAN network}

Because the CAN network has no built-in security mechanism, external interfaces have exposed it to potential attacks \cite{s20082364,Checkoway2011comprehensiveAnalysis,avatefipour2018stateoftheart}. After gaining access to the vehicle using the most common attack vectors \cite{Checkoway2011comprehensiveAnalysis}, attackers can cause several malfunction, ranging from displaying incorrect information on the vehicle dashboard to disabling vital vehicle functionality or even enabling vehicle theft\footnote{\url{https://arstechnica.com/information-technology/2023/04/crooks-are-stealing-cars-using-previously-unknown-keyless-can-injection-attacks} (last accessed: Mar. 17, 2026.)}. Such access may be achieved through physical interfaces (e.g., direct connection to the CAN bus) or by exploiting vulnerabilities in ECUs connected via wireless interfaces, such as Bluetooth, WiFi, or a 3G/4G/5G connection. 

Previous work categorizes these CAN bus attacks into three main types: \textbf{1) Fabrication}, also called message injection; \textbf{2) Suspension}; and \textbf{3) Masquerade}, or message modification attacks \cite{Cho2016}.

In a fabrication attack, an adversary injects forged CAN messages onto the bus, often at a higher frequency, in order to 
interfere with legitimate messages transmitted by safety-critical ECUs. The majority of the attacks in the literature fall into this category, including 1.1) denial-of-service (DoS) attacks, where high-priority frames monopolize the bus, 1.2) fuzzing attacks that inject random messages, and 1.3) targeted injection attacks that manipulate specific signals within a message \cite{verma_comprehensive_2020}.

In contrast, a suspension attack suppresses the transmissions of a compromised ECU, preventing essential information from being propagated to other nodes in the network.

A masquerade attack combines both strategies by first disabling the legitimate transmitter and then injecting forged messages that imitate it. Because the injected messages preserve the original ID and transmission frequency, the attack avoids message conflicts, and the overall traffic pattern appears largely unchanged, making such attacks particularly difficult to detect. Consequently, masquerade attacks are often considered timing-opaque, meaning they cannot be detected using frequency- or timing-based intrusion detection methods and instead require more advanced analysis of message payloads or physical-layer characteristics \cite{verma_comprehensive_2020}.

Although these attacks require complex technical skills, they have already been demonstrated in \cite{Cho2016}.

\subsection{Intrusion Detection Systems for CAN}

There have been many Intrusion Detection Systems (IDS) developed 
to secure the CAN bus \cite{aliwa_cyberattacks_invehicle_2021}. These IDS solutions vary in many aspects: deployment location,  network layer, used features and detection technique. To give a clear overview of the problem space, many surveys provide taxonomy and categorisation \cite{wu_survey_2019, electronics11071072, sharmin_benchmarking_2024}.

Based on the \emph{detection approach}, solutions can generally be categorized into two types:
1) signature-based methods, which rely on known attack patterns and are therefore also referred to as knowledge-based or rule-based approaches; and
2) anomaly-based methods, which model normal bus traffic and identify deviations from this baseline, and are therefore often referred to as behaviour-based approaches \cite{sharmin_benchmarking_2024}.

Based on the \emph{detection input}, we can further categorise these solutions into flow-based, payload-based, and hybrid IDSs \cite{al_jarrah_intrusion_2019, rajapaksha_ai_based_2023}.
As described in Section \ref{BC}, traffic on the CAN bus consists of CAN messages, each with its corresponding ID, and other metadata. These can be used to extract flow-based features, e.g. message frequency, message sequences, or inter-arrival time. Flow-based methods in this categorisation does not inspect the transmitted content of a message. They use extracted features to detect abnormal communication patterns without the payload.

These approaches are typically lightweight and effective at detecting attacks that disrupt normal traffic patterns, such as message flooding or injection. In contrast, payload-based IDSs inspect the content of CAN frames to identify anomalies in the transmitted data. Such methods are capable of detecting more sophisticated attacks that preserve normal timing behavior but manipulate message contents. Hybrid IDSs combine both approaches by leveraging traffic features and payload information to improve both detection accuracy and robustness against a wider range of attacks \cite{al_jarrah_intrusion_2019}.
Classifying existing works into such taxonomies are essential for a fair comparison between them. 

\subsection{IDS Methods Considered in This Study}\label{2_Methods}


In this study, to evaluate representative approaches across different CAN IDS concepts, a group of methods were selected for comparison, which were proposed in the literature.
The selected approaches capture different aspects of CAN communication, including message ordering, statistical properties of traffic, relationships between signals, and temporal behavior of vehicle states.
The following subsections briefly summarize the implemented approaches.

\paragraph{\textbf{Sequence-based flow analysis method}}

"FlowNGram", The first method compared in this study is based on the concept proposed in \cite{stabili_daga_2022}, which used $n$-gram analysis -- sequences of message IDs -- to model CAN traffic.  It constructs $n$-grams from consecutive CAN IDs to represent normal communication patterns on the bus. During operation, the observed sequences are compared with the learned model, and deviations indicate potential anomalies. This method is lightweight, which is especially important in resource-constrained automotive microcontrollers. The approach represents a typical \textit{flow-based} IDS relying on message order, and so it is expected to detect fabrication attacks, including replay, fuzzing, and denial-of-service attacks. 

\paragraph{\textbf{Machine learning–based anomaly detector}}

"MBA-OCSVM", the second method compared in this study is a machine-learning–based anomaly detection approach whose design follows the concept proposed in \cite{avatefipour_intelligent_2019}, and the main concept applies a One-Class Support Vector Machine (OCSVM) to model normal CAN traffic behavior. The model learns a decision boundary that represents normal behavior in the feature space. During detection, any observation falling outside this boundary is classified as anomalous. Because of the high complexity and nonlinearity of the data, the model parameters are finetuned using an optimization algorithm, called Modified Bat Algorithm (MBA). The approach uses statistical features, transmission frequency and message ID to build the classification model.

\paragraph{\textbf{Association-rule-based method}}

"AssocRules", the third approach compared in this study, is based on an association-rule-based anomaly detection method proposed in \cite{dangelo_association_2023}. The method focuses on modeling the vehicle state over time through correlations among CAN messages.
The approach first clusters CAN messages into multi-dimensional regions, to identify typical regions of legitimate vehicle states, and then applies association rule mining to discover frequent relationships among these states within temporal windows. During detection, deviations from the learned association rules indicate anomalous behavior.

\paragraph{\textbf{Signal-level convolutional autoencoder approach}}

"CANShield-CNN-AE", the fourth method evaluated in this study is a deep learning–based anomaly detection approach that operates at the signal level. The implementation follows the concept introduced in \cite{shahriar_canshield_2023}, where CAN payloads are decoded into individual signals, and transformed into structured two-dimensional representations that capture the temporal evolution of multiple signals. These representations are processed by convolution neural network (CNN)-based autoencoder (AE) models that learn both temporal dependencies across time and relationships between different signals. During detection, the model attempts to reconstruct the observed signal patterns, and deviations between the reconstructed and actual signals are used to identify anomalous behavior. 

\paragraph{\textbf{LSTM-based signal reconstruction method}}

"CANet-LSTM-AE", the fifth method evaluated in this study is a deep learning–based anomaly detection approach, based on the architecture proposed in \cite{hanselmann_canet_2020}, which combines recurrent neural networks with an autoencoder-based reconstruction mechanism. In this approach, separate Long Short-Term Memory (LSTM) networks are assigned to individual CAN message identifiers in order to capture the temporal dynamics of the signals associated with each identifier. The hidden states of these recurrent modules are combined into a shared latent representation describing the current state of the CAN traffic. A reconstruction network then predicts the expected values of all signals from this latent state. During detection, deviations between the reconstructed and observed signals are used to compute anomaly scores.



During the preparation of this study, we also considered including the method proposed in \cite{SAIDuCANT}. However, while reproducing the approach based on the description, we identified an inconsistency in the reported evaluation results. The confusion matrix values reported in Table IV appear to interchange the true positive and true negative counts, which results in inflated performance metrics when computed directly from the table. After recalculating the metrics using values consistent with the dataset’s attack-benign ratios, the results differed substantially from those reported in the paper. This issue was discussed with the authors, who confirmed the inconsistency. Consequently, the method was not included in the present comparison. Notably, the affected work has already been widely cited. This experience also highlights the importance of transparent and reproducible evaluation pipelines, which the proposed benchmarking framework aims to facilitate.

\section{Related Work}\label{RW}

\begin{table*}[t]
\centering
\small
\scriptsize
\renewcommand{\arraystretch}{1.15}

\caption{Comparison of related work on CAN intrusion detection system (IDS) evaluation.}
\label{tab:benchmark_comparison}

\begin{tabular}{l c c c c c c c c}
\toprule
& & \multicolumn{2}{c}{Study Scope} & \multicolumn{4}{c}{Experimental Evaluation} \\
\cmidrule(lr){3-4} \cmidrule(lr){5-9}
Study & Type & IDS Categories & Attack Types & Unified Pipeline & \# IDS Evaluated & Cross-Dataset Eval. & \# Datasets & \# Vehicle Types \\
\midrule

\cite{al_jarrah_intrusion_2019}
& SV & Broad & Broad review & N/A & N/A & N/A & N/A & N/A \\

\cite{rajapaksha_ai_based_2023}
& SV & ML-based & Broad review & N/A & N/A & N/A & N/A & N/A \\

\cite{sharmin_benchmarking_2024}
& DS & Broad & Attack dimension & Conceptual & N/A & N/A & N/A & N/A\\

\midrule

\cite{agbaje_framework_2022}
& FE & T, S, ML & F & Yes & 8 & No & 1 & 1 \\

\cite{blevins_time_based_2021}
& BE & T & F, Mq & Yes & 4 & No & 1 & 1\\

\cite{sharmin_comparative_2023}
& BE & S, ML & F, Mq & Yes & 6 & No & 1 & 1\\

\cite{pollicino_performance_2024}
& BE & T & F & Yes & 8 & No & 2 & 2\\

\midrule

\textbf{This Work}
& \textbf{FE} & T, S, ML & F, Mq & Yes & 5 & \textbf{Yes} & \textbf{7} & \textbf{7} \\

\bottomrule
\end{tabular}

\vspace{2mm}
\footnotesize

SV: survey/review paper; DS: conceptual evaluation design space; 
BE: benchmarking experiment; FE: evaluation framework with experiments.  
T: Timing-based methods; S: Statistical-based methods; ML: Machine-learning-based methods.  
F: Fabrication attacks; Mq: Masquerade attacks.

\end{table*}


In this section, we first review surveys that summarize existing CAN IDS techniques and identify key research gaps, followed by studies that specifically focus on benchmarking and comparative evaluation of CAN anomaly detection methods.

\subsection{Reviews of CAN Intrusion Detection Systems}

The review presented in \cite{al_jarrah_intrusion_2019} categorizes 42 IDS approaches for intra-vehicle networks and outlines the key foundational techniques that have shaped subsequent research in the field.
From their analysis, 86\% of the works focused on fabrication attacks, and only 7\% targeted masquerade attacks, which showed a significant research gap at the time. Later works, like \cite{rajapaksha_ai_based_2023} shows that many current work now targets the more complex masquerade attack type, which involved a shift to ML-based approaches.

Yet, as \cite{al_jarrah_intrusion_2019} points out, likely no single solution could provide a full detection capability of all attacks.
Rule-based IDSs are often simple and fast when a fixed rule set is used, but they suffer from limited generalization capability and strong dependence on prior knowledge. Time and frequency analysis methods are suitable for detecting attacks that affect message timing or transmission frequency, however, they are unable to detect payload modification attacks. Computational intelligence and information theory–based approaches generalize better and do not require explicit prior knowledge, but they are more complex and dependent on data availability.

Thus the design process of a proper IDS solution should consider  methods with different capabilities, to optimize resources as well as generalize to unknown attacks. This process require the fair and interpretable comparison of state-of-the-art techniques, benchmark models and benchmark datasets.

However, the review in \cite{al_jarrah_intrusion_2019} highlighted a significant shortage of publicly available benchmark datasets containing real intra-vehicle data, which results in evaluation outcomes that are often not directly comparable across studies. Since then, two new benchmark datasets have been introduced \cite{gazdag_crysys_2023, verma_comprehensive_2020}. These datasets are recommended due to their inclusion of multiple real attack types together with benign traffic collected under various driving conditions \cite{rajapaksha_ai_based_2023}. Nevertheless, many studies continue to rely on a wide range of datasets, including proprietary or non-public data sources. Consequently, reported performance metrics remain difficult to compare across different works, and benchmark datasets have not yet been consistently adopted in current research.
Another major research gap identified in \cite{al_jarrah_intrusion_2019} is the absence of a standard benchmark detection model against which newly proposed approaches can be evaluated. This issue persists today.

\newpage

A more recent survey on AI-based IDSs reviewed 102 studies between 2016 and 2022, and summaries the most common detection methods, evaluation datasets, attack types, and performance evaluation \cite{rajapaksha_ai_based_2023}. 
They also introduce an AI-based IDS taxonomy, and review benchmark datasets available to train and evaluate AI-based methods.
Their taxonomy supports the development of a new AI-based IDS, as the selected features should indicate a concise decision in the targeted attack types. They note that an ID-based approach could generalize better to other vehicles than a payload-based variant, as the CAN payload is extremely unique, and despite the functionality of the IDs is different, frequent or sequential behavior could be common.
From their findings, they argue that deep learning models have achieved better accuracy than traditional ML models, but also have high resource requirements which is often a concern for IVN devices. They suggest that hybrid models and ensemble models could increase the detection power, as these models can improve performance while decreasing the weakness of individual models.


\subsection{Benchmarking Surveys on CAN Anomaly Detection}


The comparison framework, presented in \cite{agbaje_framework_2022}, was one of the first to facilitate consistent, fair comparisons of proposed CAN IDS approaches. Their work uses the OTIDS dataset \cite{lee_otids_2017}. Their framework preprocesses the dataset for timing-based, statistical and machine learning methods, and separates some of the attacked logs as training data for signature-based approaches. Notably, they also implement and compare some of the common detection approaches, and based on their results, they suggest that methods considering interrelationship between messages (ML models and graph-based models), are better suited for the task.

One of the recent works on benchmarking surveys, \cite{sharmin_benchmarking_2024} proposes a higher level approach to the problem, an evaluation design space. Unlike prior surveys that primarily categorise IDS approaches, their work focuses on the methodology of the benchmarking process in this domain.
Their design space enumerates the various decision points for these aspects, and aims to guide future CAN IDS evaluation experiments, highlighting how experimental design choices influence the interpretation and comparability of reported results.

An empirical comparison of time-based anomaly detectors is presented in \cite{pollicino_performance_2024}. The authors evaluate eight timing-based CAN IDS methods, as they represent one of the most analyzed group of methods, on two datasets, including a newly collected dataset called Ventus dataset, and the publicly available OTIDS dataset \cite{lee_otids_2017}.
The study also points out that only a limited number of state-of-the-art implementations are publicly available, which makes fair comparison with newly proposed methods difficult.

Another comparative evaluation is presented in \cite{sharmin_comparative_2023}, where a variety of anomaly-based IDS methods were tested on the ROAD dataset \cite{verma_comprehensive_2020}. In addition to conventional security metrics, the authors report balanced accuracy, informedness, markedness, and Matthews correlation coefficient (MCC), arguing that these metrics are more suitable for imbalanced datasets \cite{chicco_matthews_2021}, which are often the case in CAN intrusion detection.

Another benchmarking study focusing on timing-based CAN intrusion detection methods is presented in \cite{blevins_time_based_2021}. The authors compare four anomaly detection approaches using the ROAD dataset.
Their results show that simple distribution-agnostic methods could outperform distribution-based approaches, but despite the strong detection performance, even detectors with very high precision can generate a large number of alerts in practice, which also reinforces the need for evaluation metrics suited to imbalanced datasets.

As Table~\ref{tab:benchmark_comparison}
shows, while these works provide comparative benchmarking of detection methods, experiments are typically conducted on a single dataset. Cross-dataset evaluation, which would allow assessing whether IDS performance generalizes across different vehicles, attack implementations, and data collection environments, has received little attention in the literature. Consequently, the reported performance results are often strongly tied to the specific dataset and experimental setup used in each study.
In contrast, the proposed framework in this work enables consistent cross-dataset evaluation across multiple IDS categories and multiple publicly available CAN intrusion datasets. In this work, we run our experiments on five different IDS method on seven different datasets.


\section{Benchmarking Framework}\label{BF}

To address the comparability challenges identified in the previous section, we propose a benchmarking framework designed for systematic evaluation of CAN IDSs.




The evaluation is guided by the following research questions:

\begin{itemize}
\item \textbf{RQ1:} How much does IDS detection performance vary across different CAN intrusion datasets?

\item \textbf{RQ2:} If performance varies across datasets, do the relative rankings of IDS methods remain stable across datasets with different traffic characteristics and attack implementations?

\item \textbf{RQ3:} Assuming these variations, can performance results obtained on a single dataset reliably predict IDS performance on other datasets and vehicle environments?
\end{itemize}


\subsection{Design Goals and Requirements}\label{4_DGR}

Designing a benchmarking framework for CAN intrusion detection systems presents several challenges due to the heterogeneity of available datasets and detection methods. In order to enable meaningful comparison between IDS approaches, these challenges must be considered.

\paragraph{\emph{\textbf{Challenge 1: Heterogeneous dataset representations}}}

Although the CAN protocol defines a well-structured message format, publicly available datasets store CAN traffic in a variety of representations:
\emph{Raw CAN logs} are typically recorded using tools such as \texttt{candump} and may contain classical CAN 2.0 frames or CAN~FD (CAN Flexible Data-rate) frames. \emph{Structured tabular datasets} are commonly provided as CSV files, where CAN frames are 
often preprocessed. \emph{Signal-level data} provide higher-level representation and decode CAN payloads into individual signals rather than storing them as raw frames. \emph{Synthetic datasets} are often generated directly in structured tabular form without the corresponding raw network traces.


\paragraph{\emph{\textbf{Challenge 2: Heterogeneous detection outputs}}}

Detection methods differ both in their output representation and in the granularity at which they analyze CAN traffic. Some approaches perform \emph{binary classification}, directly labeling messages as benign or malicious. Others produce \emph{continuous anomaly scores} that require threshold selection to determine detections. Certain methods operate at the \emph{message level}, while others analyze sequences of messages using \emph{time-based aggregation or sliding windows}. Apart from the difficulty of comparing different output formats—for example, evaluation at the window level can artificially inflate perceived performance because multiple CAN messages are aggregated into a single decision—this heterogeneity also affects the applicability of certain IDS approaches. For example, when datasets provide only decoded signal-level representations, the original CAN frames are no longer available, making it difficult to apply methods that rely on the raw CAN frames. These differences make direct comparison difficult, since evaluation metrics and detection thresholds may depend on the type of output produced by the IDS.

\paragraph{\emph{\textbf{Challenge 3: Heterogeneous attack scenarios}}}

Not all IDS methods are designed to detect the same types of attacks.
It would be misleading to directly compare methods designed for fundamentally different threat models, for example, approaches targeting fabrication attacks with those specifically designed to detect masquerade attacks. At the same time, IDS methods that target a particular class of attacks should demonstrate robustness across variations of that attack family. For instance, methods primarily evaluated on DoS attacks should ideally also detect related fabrication attacks such as fuzzy message injection, as these attacks similarly violate the normal message transmission pattern of the network.


To address these challenges, the proposed framework introduces three key components:

\begin{enumerate}
    \item a unified dataset representation that normalizes heterogeneous CAN datasets,
    \item a common interface for IDS methods that enables integration of different detection approaches, and
    \item a uniform evaluation protocol that ensures fair comparison between heterogeneous detection outputs.
\end{enumerate}

In the proposed framework,
binary detectors are interpreted as fixed-threshold decisions, while score-based detectors are evaluated across varying thresholds.
Performance evaluation relies primarily on threshold-independent metrics.
Receiver operating characteristic (ROC) curves and the corresponding area under the curve (ROC-AUC) are computed for all scoring-based detectors.

Since CAN datasets are typically highly imbalanced,
we primarily report \textit{balanced accuracy} ($bACC$), \textit{F1-score} ($F1$), and the \textit{Matthews correlation coefficient} (MCC), which are commonly used evaluation metrics in CAN IDS benchmarking studies \cite{sharmin_comparative_2023}. Let $TP$, $TN$, $FP$, and $FN$ denote the entries of the confusion matrix, and $TPR = \tfrac{TP}{TP+FN}$ and $TNR = \tfrac{TN}{TN+FP}$.

\begin{equation}
\label{eq:metrics}
\begin{aligned}
bACC &= \frac{TPR + TNR}{2} \\\\[3pt]
F1 &= \frac{2TP}{2TP + FP + FN} \\\\[3pt]
MCC &= \tfrac{TP\cdot TN - FP\cdot FN}
{\sqrt{(TP+FP)(TP+FN)(TN+FP)(TN+FN)}}\\[3pt]
\end{aligned}
\end{equation}



\subsection{Datasets}\label{4_DS}
\begin{figure*}[ht]
  \centering
  \begin{subfigure}{\textwidth}
    \includegraphics[width=\linewidth]{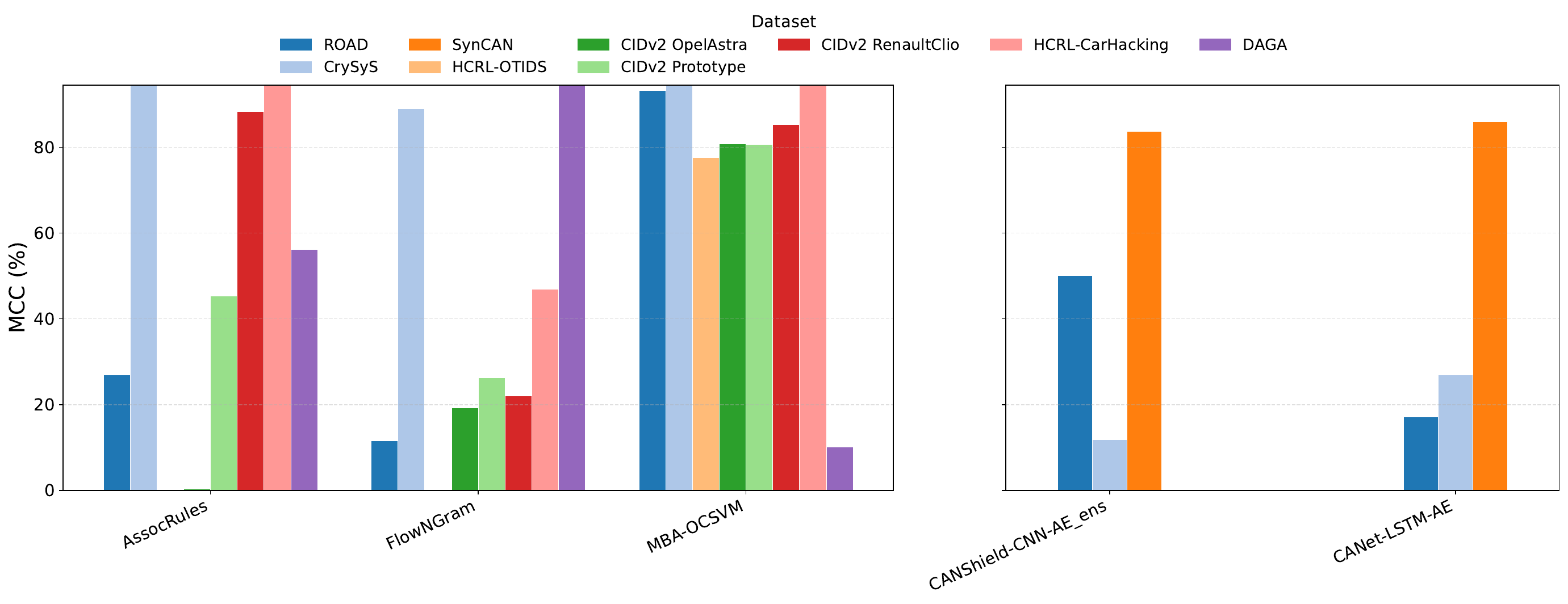}
  \end{subfigure}
  
  \caption{Matthews Correlation Coefficient (MCC) of methods across datasets.}
  \label{fig:pooled-best-metrics}
\end{figure*}


To investigate whether results reported in different studies are directly comparable, our framework evaluates IDS methods across multiple publicly available CAN intrusion detection datasets.


\paragraph{\textbf{OTIDS Dataset}}

The CAN Intrusion Dataset released by the Hacking and Countermeasure Research Lab (HCRL) introduced in \cite{lee_otids_2017}, is one of the earliest publicly available datasets for CAN intrusion detection research. The dataset contains real CAN traffic collected from a Kia Soul vehicle and includes injected attack scenarios.
Due to its early release, OTIDS has been widely adopted in IDS research.

Subsequent analysis of the dataset revealed several limitations that may affect its suitability for general benchmarking. In particular, inconsistencies between the dataset documentation and the actual attack intervals have been reported, and the impersonation attack appears to correspond to a fabrication attack rather than a true masquerade attack, since legitimate messages are not fully suppressed~\cite{verma_comprehensive_2020}. 

\paragraph{\textbf{DAGA Dataset}}

The dataset provided in \cite{stabili_daga_2022} consists of real CAN traffic collected from a 2016 Volvo V40 Kinetic model, from different driving scenarios. The dataset has fabrication attack scenarios, including replay attacks (Single ID Replay, Ordered Sequence Replay, Arbitrary Sequence Replay), fuzzing attacks, and DoS attacks implemented through message injection on the CAN bus.

\paragraph{\textbf{Automotive CAN Bus Intrusion Dataset}}
The Automotive Controller Area Network (CAN) Bus Intrusion Dataset v2~\cite{cidv2} is one of the few publicly available CAN intrusion datasets that includes traffic collected from multiple vehicle platforms. The dataset contains CAN logs from two real vehicles (an Opel Astra and a Renault Clio) as well as from a laboratory CAN bus testbed. Therefore, we report metrics separately for the two vehicles and the testbed, however, for clarity, we treat them as a single dataset when reporting the total number of datasets used.

An important advantage of this dataset is that it includes attack scenarios rarely available in public CAN datasets, such as diagnostic protocol attacks and suspension attacks on real traffic~\cite{verma_comprehensive_2020}. In addition, the same attack types are available across three distinct CAN environments, enabling comparative cross-vehicle evaluation.

However, most attack traces were generated through post-processing of recorded CAN logs by adding, modifying, or removing frames and adjusting timestamps accordingly~\cite{verma_comprehensive_2020}. Because CAN message timing is strongly influenced by arbitration and message priorities, such timestamp modifications may reduce the realism of the resulting traffic traces.

\paragraph{\textbf{Car-Hacking Dataset}}
The Car-Hacking Dataset, released by HCRL, contains CAN traffic collected from a Hyundai YF Sonata model while message injection attacks were actively performed \cite{cidv2}.
The dataset includes both normal driving traffic and fabricating attack scenarios, including DoS, fuzzing attacks, and spoofing attacks targeting specific vehicle functions such as the drive gear and engine RPM signals.
This dataset seems to be the most widely used source of CAN data in the literature. However, similar to many other datasets, the attacks consist primarily of message injection scenarios.

Limitations reported regarding this dataset are that large gaps in message timestamps appear at the end of attack intervals, likely caused by the vehicle entering a standby state after the injection process stops~\cite{verma_comprehensive_2020}. Additionally, the attacks were conducted while the vehicle was stationary, whereas the ambient traffic was recorded during driving, resulting in differences between training and testing data distributions.

\paragraph{\textbf{SynCAN Dataset}}
The SynCAN dataset \cite{hanselmann_canet_2020} consists of generated signal-level CAN traffic derived from decoded CAN payloads.
Since the dataset consists of synthetic signal data rather than measurements collected from a real vehicle, its traffic characteristics may differ from real CAN traffic, and the lack of the original raw CAN frames limits the applicability of the dataset to IDS methods that operate directly on frame-level CAN data \cite{verma_comprehensive_2020}.

\paragraph{\textbf{ROAD Dataset}}

The Real ORNL Automotive Dynamometer (ROAD) dataset, introduced in \cite{verma_comprehensive_2020}, was designed to address limitations of earlier CAN IDS datasets by providing more diverse and realistic attack scenarios;
fuzzing, targeted message fabrication and masquerade attacks: 1) correlated signal attacks, where four different false wheel speed values were used; 2) max speedometer attack, where the speedometer was forced to falsely display a maximum value; 3) max engine coolant temperature attack, forcing the engine coolant signal to be the maximum; 4) reverse light attack, where the state of the reverse lights were forced.
For all these attack types, the dataset provide a version where, in post-processing, original target frames were removed to simulate a masquerade attack.
An important advantage is that the dataset provides both raw CAN messages and signal-translated versions of the data, with an obfuscated DBC file.

Although the attacks were executed on a real vehicle rather than a laboratory testbed, the experiments were conducted on a dynamometer for safety reasons. While this setup ensures that attacks are performed on an authentic in-vehicle CAN network, the vehicle operates in a controlled testing environment, and therefore external conditions present during real-world driving scenarios are not reflected in the data~\cite{gazdag_crysys_2023}.

Since publication, this dataset was used in multiple benchmark studies \cite{blevins_time_based_2021, sharmin_comparative_2023}.

\paragraph{\textbf{CrySyS CAN Dataset}}

Similar to the ROAD dataset, the CrySyS CAN dataset introduced in~\cite{gazdag_crysys_2023} was created to address the lack of publicly available and sufficiently comprehensive benchmark datasets. Both datasets provides realistic CAN traffic traces and diverse attack scenarios.

The dataset contains more than 2.5 hours of benign CAN traffic recorded from a real vehicle in various driving scenarios, including urban driving, country roads, and motorway traffic. In total, 26 benign traces were captured, and both fabrication and masquerade attacks are present. The attacks target vehicle speed and engine revolution signals and are implemented using several signal modification strategies, including constant value replacement, replayed signal values, and incremental signal offsets. In addition to attacks affecting a single signal, the dataset also includes scenarios where two signals are manipulated simultaneously. The dataset also provides tools for generating additional attack scenarios, allowing researchers to extend the dataset for further experimentation. This dataset was also used in CAN IDS development studies\cite{koltai_supporting_2024}.


\section{Evaluation}\label{E}

\begin{table*}[t]
\centering
\caption{Balanced Accuracy / F1-score / MCC (\%) for three methods across datasets. Best overall method per dataset is shown in bold when it achieves the strongest results across the three reported metrics.}
\label{tab:multi_metric_results}
\setlength{\tabcolsep}{5pt}
\renewcommand{\arraystretch}{1.1}
\begin{tabular}{lccc}
\toprule
Dataset & AssocRules & FlowNGram & MBA-OCSVM \\
\midrule
HCRL-OTIDS         & 50.00 / 79.58 / 0.00   & 50.00 / 57.95 / 0.00   & \textbf{89.04 / 86.91 / 77.59} \\
CIDv2 Opel Astra  & 50.01 / 30.39 / 0.24   & 80.02 / 30.07 / 19.18  & \textbf{82.78 / 79.20 / 80.80} \\
CIDv2 Prototype   & 72.69 / 64.06 / 45.27  & 66.36 / 43.77 / 26.30  & \textbf{85.01 / 82.36 / 80.64} \\
CIDv2 Renault Clio& \textbf{90.00 / 88.89 / 88.43} & 70.97 / 23.08 / 22.02  & 87.17 / 85.28 / 85.34 \\
HCRL-CarHacking       & \textbf{98.82 / 98.81 / 97.90} & 82.75 / 69.24 / 46.94  & 98.26 / 98.44 / 96.28 \\
CrySyS            & \textbf{97.30 / 97.23 / 96.69} & 98.31 / 90.13 / 89.08  & 95.38 / 94.63 / 94.43 \\
DAGA              & 67.60 / 94.66 / 56.24  & \textbf{99.81 / 99.93 / 99.58} & 56.59 / 97.82 / 10.05 \\
ROAD              & 57.75 / 68.12 / 26.96  & 55.60 / 63.30 / 11.61  & \textbf{96.49 / 96.32 / 93.28} \\

\toprule
Dataset & CANet-LSTM-AE & CANShield-CNN-AE  \\
\midrule
SynCAN         & 90.95 / \textbf{87.79} / \textbf{85.92}   & \textbf{91.46} / 86.33 / 83.68   \\
ROAD         & 69.65 / 15.57 / 17.18   & \textbf{76.24} / \textbf{65.69} / \textbf{50.15}   \\
CrySyS         & \textbf{66.04} / \textbf{30.83} / \textbf{26.96}   & 58.83 / 26.53 / 11.78   \\

\bottomrule
\end{tabular}
\end{table*}

This section evaluates the proposed benchmarking framework and investigates whether IDSs exhibit consistent performance across different CAN datasets. 
The experiments assess whether methods reported in the literature remain comparable when evaluated under identical conditions within a unified benchmarking framework.


All experiments were conducted using the proposed benchmarking framework implemented in Python~3.12, using identical training and testing policies, training exclusively on benign CAN traffic. To ensure reproducibility, the complete framework source code -- including the method implementations described in Section~\ref{2_Methods} and adapters for the dataset presented in Section~\ref{4_DS} -- is publicly available on GitHub\footnote{Code available: \href{https://github.com/CrySyS/Cross-Dataset-Study-of-Automotive-IDS-Evaluation}{https://github.com/CrySyS/Cross-Dataset-Study-of-Automotive-IDS-Evaluation}}.



The framework reports evaluation results at two granularities: 1) message-level metrics, and 2) window-level metrics computed on fixed windows using the canonical labeling rule, where a window is labeled as an attack if it contains at least one malicious message (see Appendix~\ref{appendix:window} for a discussion on window size selection).
For scoring-based methods, detection thresholds are selected using best-F1 operating point. Confusion-matrix-based metrics are then computed consistently. For methods producing binary outputs, the native predictions are evaluated directly.
Given the strong class imbalance in CAN IDS datasets, the analysis focuses on imbalance-robust metrics (Equation~\eqref{eq:metrics}). In addition, per-attack-family analyses are performed to avoid over-interpreting aggregate performance metrics.

Some factors may influence the interpretation of the presented results.
With respect to internal validity, the evaluated IDS implementations rely on default hyperparameter settings and configurations derived from available reference implementations or descriptions in the original publications. Although this approach ensures consistency across experiments, alternative hyperparameter tuning strategies could affect the absolute performance values reported in this study.
Where applicable, we adjusted the evident parameters and performed the minimal modifications required to apply the methods to the evaluated datasets. However, we deliberately avoided extensive tuning or dataset-specific adaptations. From a practical deployment perspective, an IDS solution should remain effective across different CAN environments without requiring substantial reconfiguration. While minor adjustments related to CAN bus specifics (e.g., identifier sets) can reasonably be handled by integration layers or third-party tooling, approaches that require extensive modifications or repeated manual tuning for each new environment would be difficult to deploy in practice.

\subsection{Cross-dataset comparison}

Table~\ref{tab:multi_metric_results} summarize the cross-dataset detection performance of the evaluated IDS methods using the best available evaluation level, and Figure~\ref{fig:pooled-best-metrics} shows the MCC metrics.
See Appendix~\ref{appendix:datasets} for a detailed description of the method–dataset evaluation setup.
The results reveal substantial changes in method ranking across datasets. No single method consistently achieves the best performance across all datasets, and the same method can perform very differently depending on the dataset.

Notably, each method achieves strong performance on the datasets used in its original evaluation. However, the ranking, that is, the relative performance ordering of the methods changes considerably when they are evaluated on other datasets. For example, FlowNGram achieves near-perfect performance on the DAGA dataset, reaching a balanced accuracy of 99.81\% and an F1 score of 99.93\%. Similarly, MBA-OCSVM and AssocRules achieve good results on the HCRL-CarHacking dataset, which was also used in their original evaluations.

When evaluated on other datasets, however, the performance of these methods in some cases decreases significantly. For instance, FlowNGram performs substantially worse on CIDv2 and HCRL-OTIDS, while MBA-OCSVM shows a notable performance drop on the DAGA dataset.
This suggests that methods relying on flow-level statistical features may be sensitive to differences in dataset structure and attack generation methodology.



In contrast to the previously discussed methods, CANet-LSTM-AE and CANShield-CNN-AE operate on signal-level representations, however, their performance also varies notably across datasets. On the SynCAN dataset, both methods achieve strong performance, with CANShield-CNN-AE slightly outperforming CANet-LSTM-AE in terms of balanced accuracy. 
However, their performance diverges significantly on the ROAD dataset, where CANShield-CNN-AE maintains relatively strong detection capability, CANet-LSTM-AE exhibits a notable degradation. On the CrySyS dataset both methods achieve comparatively lower performance overall.

To better understand the observed performance differences, we further analyzed the characteristics of the underlying datasets. In particular, the SynCAN dataset is synthetically generated and contains a small number of signals with relatively homogeneous temporal behavior. In contrast, the ROAD dataset is derived from real vehicle traces and contains a substantially larger and more diverse set of signals with markedly different statistical properties. Although a subset of relevant signals was selected from ROAD for training, the remaining variability in signal characteristics is still significantly higher than in SynCAN.
We hypothesize that this difference contributes to the observed performance degradation. While CANet employs separate LSTM models per signal, these representations are subsequently combined into a shared latent space through fully connected layers. This design may be well-suited for datasets with relatively homogeneous signals, but may become less effective when modeling highly heterogeneous signal distributions, as in real-world datasets such as ROAD. Although the SynCAN dataset is widely adopted in the literature, its simplified structure may not fully capture the complexity of real CAN traffic, potentially leading to overly optimistic performance estimates.
Finally, we performed a grid search over key hyperparameters suggested in the original works and commonly used in practice, and while minor improvements were observed, the overall performance trends remained unchanged (see Appendix~\ref{appendix:grid_search}). This suggests that the performance gap is not solely due to suboptimal parameter settings, but may instead require more substantial architectural adaptation. At the same time, such adaptations are not straightforward and raise practical questions about how these methods should be systematically tuned for deployment in new environments.

\subsection{Per-attack and per-vehicle analysis}

As discussed previously, not all IDS methods are designed to detect the same types of attacks, but at the same time, IDS methods that target a particular class of attacks should demonstrate robustness across variations of that attack family.
Given that the datasets used in this work contain different implementations of similar attack concepts, evaluating methods only on a per-dataset basis may obscure this cross-variant robustness. Therefore, in addition to reporting results separately for each dataset, we also aggregate subresults across datasets according to seven normalized attack families. The grouping of the different attack implementations into these families is shown in Table~\ref{tab:per_attack_type}. This allows us to analyze whether a detection method performs consistently on the same type of attack across different datasets, despite variations in implementation details and environments.

For the signal-level methods (CANet-LSTM-AE and CANShield-CNN-AE), a comparable breakdown is not provided. The message modification attacks used to evaluate these approaches are significantly more diverse in their implementation, although they share the common principle of altering the values of targeted signals, so grouping these attacks into unified families would be less meaningful and could lead to misleading interpretations.

The results show that detection performance varies substantially across attack types. High detection performance is typically achieved for high-volume injection attacks such as DoS. For example, AssocRules method achieves an average F1 score of 86.1\% for DoS attacks, and MBA-OCSVM scores 94.2\%.

A notable observation concerns targeted ID injection and signal manipulation attacks. While MBA-OCSVM performs well in the case of targeted ID injection, its performance drops significantly for signal manipulation attacks. In contrast, AssocRules maintains a more consistent performance across both attack types, likely due to its hybrid design, which incorporates payload content in addition to message identifiers.
In contrast, more subtle attacks -- such as suspension and replay attacks -- are considerably harder to detect, and detection performance for these drops substantially.

These results suggest that IDS methods often learn dataset-specific patterns of normal CAN traffic, which may limit their ability to generalize when detecting anomalous behavior across different datasets.
Because the datasets vary in several dimensions simultaneously, it is challenging to isolate the effect of individual factors. The observed differences therefore likely arise from a combination of dataset-specific properties. In the case of the CIDv2 datasets, however, the attack implementations are identical, which suggests that the observed differences are more likely related to variations in normal traffic patterns rather than differences in the attacks themselves.

These observations suggest that a single IDS method may not be sufficient to reliably detect the full spectrum of attack types. Instead of focusing solely on optimizing the performance of a method on a particular dataset, effective IDS solutions should aim to combine complementary detection approaches that collectively cover a broader range of attack behaviors across multiple datasets.
This includes identifying which attack types are addressed by each method and determining how multiple methods can be combined to achieve comprehensive protection. Ideally, such combinations should aim to minimize resource usage while maximizing coverage, potentially prioritizing lightweight or computationally inexpensive techniques when multiple methods provide similar detection capabilities.

Table~\ref{tab:multi_metric_results} also reports per-vehicle results for the CIDv2 dataset. While some of the remaining datasets contain real CAN traffic collected from real vehicles, each dataset typically represents a single vehicle and therefore does not enable direct cross-vehicle comparison within the same dataset.  In contrast, CIDv2 applies the exact same attack implementations to two different CAN environments. This controlled setup allows us to attribute differences in detection performance to variations in the underlying CAN environments rather than to differences in the attack implementations themselves.

Results indicate that performance varies across different vehicle platforms as well. For example, the AssocRules detector achieves F1 score of 30\% on the Opel Astra subset, while performance improves to 89\% on the Renault Clio subset. Similarly, MBA-OCSVM achieves better F1 score on the Renault Clio subset, while showing slightly lower performance on the Opel Astra dataset. In contrast, FlowNGram performs consistently worse across all CIDv2 vehicle variants. 

\newpage
These results suggest that sequence-based detection methods may be particularly sensitive to differences in message ordering and traffic characteristics between vehicles.

\begin{table*}[ht]
\centering
\caption{Best F1 Score (\%) per Method and Attack Family}
\label{tab:per_attack_type}
\begin{tabular}{lccccccc}
\toprule
\textbf{Method} & \textbf{Diagnostic} & \textbf{DoS} & \textbf{Fuzzy} & \textbf{Replay} & \textbf{Suspension} & \textbf{Targeted ID Injection} & \textbf{Targeted Signal Manipulation} \\
\midrule
AssocRules   & \textbf{53.2} & 86.1 & \textbf{79.0} & \textbf{50.7} & \textbf{13.5} & 65.2 & \textbf{69.7} \\
FlowNGram    & 20.4 & 40.7 & 14.6 & 26.3 & 3.9  & 29.0 & 36.5 \\
MBA-OCSVM    & 50.3 & \textbf{94.2} & 43.1 & 2.7  & 6.2  & \textbf{90.4} & 33.8 \\
\bottomrule
\end{tabular}
\end{table*}


\section{Conclusion}



This paper presented a unified benchmarking framework for the systematic evaluation of CAN intrusion detection systems across heterogeneous datasets and vehicle types.
By introducing a unified dataset representation, a common IDS interface, and a standardized evaluation pipeline, the framework enables reproducible and fair comparison of IDS methods under identical experimental conditions, allowing us to assess whether a method that performs well on one dataset remains reliable under different conditions.

Using this framework, we evaluated five representative IDS approaches across seven publicly available CAN intrusion datasets; including two vehicle environments with identical attack implementations; and seven aggregated attack families across the different datasets. The results reveal that IDS performance is strongly influenced by dataset characteristics, attack implementations, and vehicle environments.
The results provide empirical insights into how dataset characteristics influence the reported performance of CAN IDS methods.

\textbf{RQ1:} IDS detection performance varies substantially across publicly available datasets. Methods that achieve near-perfect performance on certain datasets may exhibit significantly lower balanced accuracy or F1-scores on others. These variations indicate that dataset-specific properties, including attack generation strategies, background traffic patterns, and class imbalance, strongly influence detection outcomes, even within the same general attack category, such as fabrication attacks or masquerade attacks.

\textbf{RQ2:} The relative ranking of IDS methods can change considerably across publicly available datasets. We observed that evaluated methods achieve strong performance on some datasets while clearly underperforming on others. Consequently, method comparisons reported in different studies may lead to conflicting conclusions when evaluated on different datasets.

\textbf{RQ3:} Performance results obtained on a single dataset do not reliably predict IDS performance on other datasets or vehicle environments. Even under the same attack implementations, substantial cross-dataset and cross-vehicle performance variation was observed.

Taken together, these findings suggest that comparing only reported results from different studies with different datasets, should be interpreted with caution. Single-dataset evaluations may overestimate the generalization capability of IDS methods. 
These observations are consistent with the broader machine learning literature on spurious correlations, where models may rely on dataset-specific artifacts rather than the underlying phenomenon of interest. In the context of CAN intrusion detection, such artifacts can arise when attacks are recorded under specific conditions, for example, when the vehicle is stationary, or when traffic is collected on a laboratory test bench rather than during real driving. As a result, models trained and evaluated on a single dataset may implicitly learn these dataset-specific characteristics instead of generalizable attack indicators. Cross-dataset benchmarking therefore provides a more reliable way to assess the robustness and reliability of IDS methods under varying operational conditions.


More broadly, these results suggest that CAN intrusion detection should not be evaluated as isolated machine learning models but as components of a larger security architecture. System-level approaches that integrate complementary detectors are likely to provide more robust and generalizable solution than individual detectors optimized for a particular dataset.

To support this perspective, the benchmarking framework proposed in this work enables systematic cross-dataset evaluation and provides a more reliable basis for comparing CAN IDSs. In practice, benchmark reports should include multi-dataset results on different vehicles, per-attack-family breakdowns, and imbalance-aware metrics.

Instead of searching for a single universal IDS, future research should focus on designing modular detection architectures that combine lightweight detectors with complementary capabilities. Such architectures are also more cost-efficient, as lightweight detectors can filter out a large portion of simple attacks early in a detection pipeline, so only a reduced subset of traffic needs to be processed by more computationally expensive detectors.
Additionally, standardized reporting of runtime performance and resource consumption should be investigated to better assess the deployability of IDSs in resource-constrained automotive systems.
While this work focuses on robustness with respect to dataset and environment variation, another important research direction concerns adversarial robustness. In practical deployments, attackers may adapt their behavior to evade detection mechanisms once the IDS design becomes known. Systematic evaluation of CAN IDSs under adaptive adversaries therefore remains an important open research problem that has received limited attention so far.

\section*{\uppercase{Acknowledgment}}
This work was supported by the National Research, Development and Innovation Fund of Hungary (NKKP-ADVANCED\_25-153325).

\newpage

\bibliographystyle{IEEEtran}
\bibliography{bibliography}

\appendices

\section{Window Size Selection}
\label{appendix:window}

\begin{figure*}[ht]
  \centering
  \begin{subfigure}{0.48\textwidth}
    \centering
    \includegraphics[width=\linewidth]{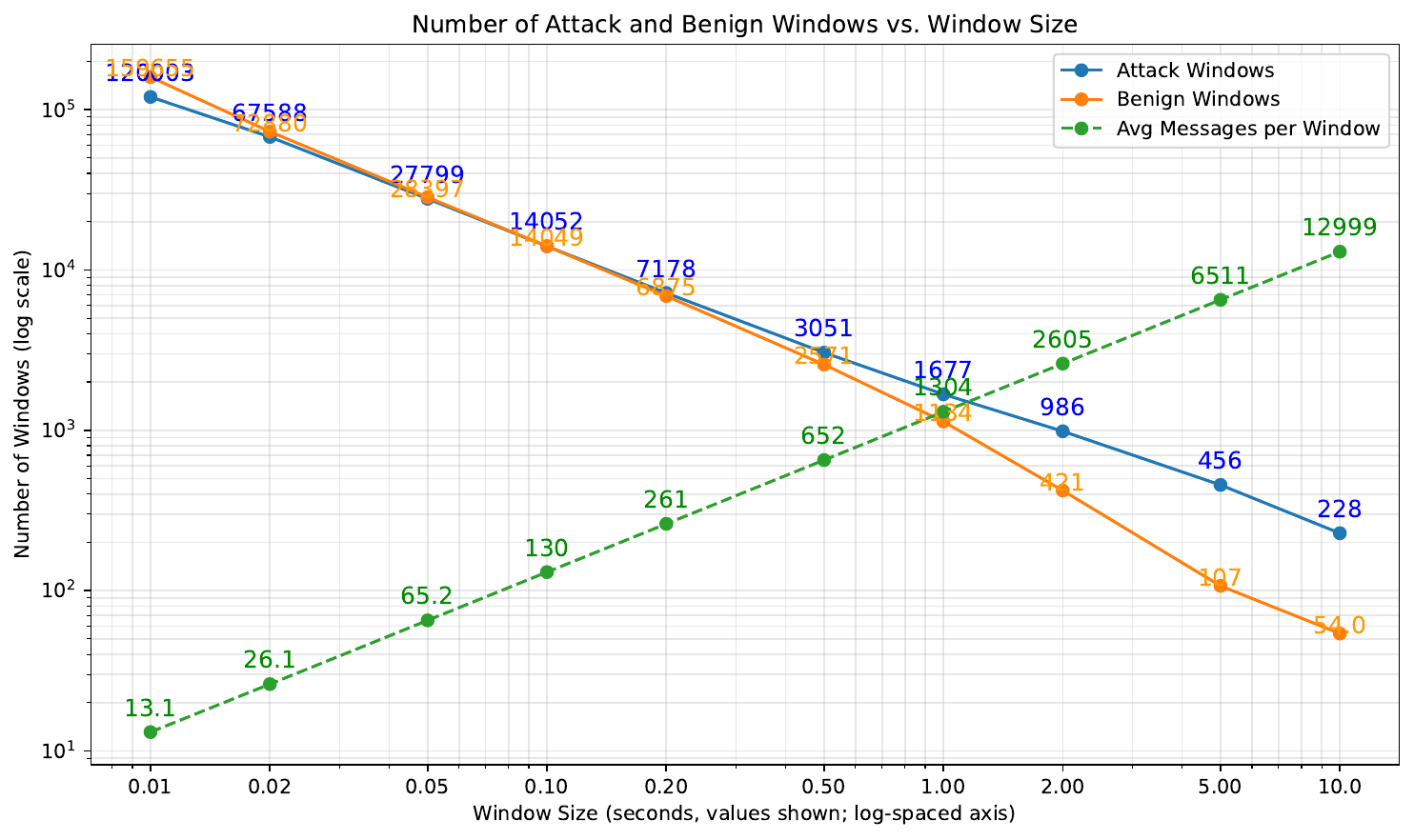}
    \caption{HCRL-CarHacking dataset (DoS attack).}
  \end{subfigure}
  \hfill
  \begin{subfigure}{0.48\textwidth}
    \centering
    \includegraphics[width=\linewidth]{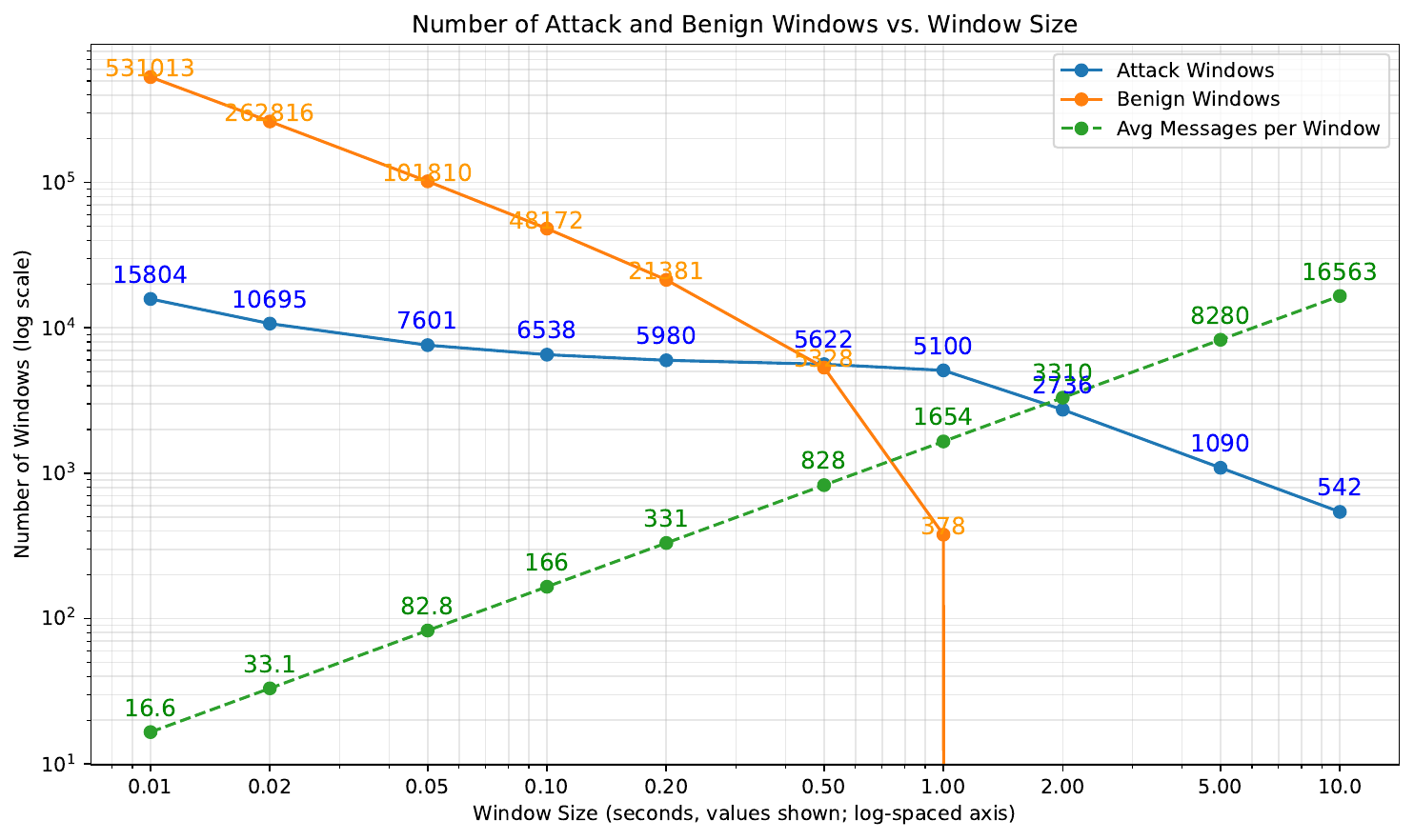}
    \caption{DAGA dataset (DoS attack).}
  \end{subfigure}

  \caption{
  Effect of window size on dataset composition.  }
  \label{fig:window-analysis}
\end{figure*}

\begin{table*}[!b]
\centering
\caption{Grid search results for the CANet-LSTM-AE method on the ROAD and CrySyS dataset.}
\label{tab:appendix_canet_grid_search}
\begin{tabular}{lllllllll}
\toprule
\textbf{Dataset}     & \textbf{Best epoch} & \textbf{Learning-rate}       & \textbf{Hidden size} & \textbf{Quantile} & \textbf{TPR(\%)} & \textbf{TNR(\%)} & \textbf{F1(\%)}   & \textbf{AUC(\%)} \\
\midrule
\multirow{12}{*}{ROAD}       & 27   & 3.00E-04 & 5            & 0.95     & 47.8 & 74.1 & 8.8 & 67.3   \\
 & 27          & 3.00E-04 & 5            & 0.99     & 46.2 & 76.6 & 7.7 & 67.3   \\
 & 27          & 3.00E-04 & 5            & 0.995    & 46.2 & 77.1 & 7.7 & 67.3   \\
 & 27          & 3.00E-04 & 5            & 0.999    & 46.2 & 77.7 & 7.7 & 67.3   \\
 & 29          & 5.00E-05 & 5            & 0.95     & 46.2 & 76.7 & 7.7 & 64.5   \\
 & 29          & 5.00E-05 & 5            & 0.99     & 46.2 & 77.8 & 7.7 & 64.5   \\
 & 29          & 5.00E-05 & 5            & 0.995    & 46.2 & 77.8 & 7.7 & 64.5   \\
 & 29          & 5.00E-05 & 5            & 0.999    & 46.2 & 77.8 & 7.7 & 64.5   \\
 & 28          & 1.00E-04 & 8            & 0.95     & 47.5 & 73.0 & 8.2 & 61.9   \\
 & 28          & 1.00E-04 & 8            & 0.99     & 46.2 & 76.6 & 7.6 & 61.9   \\
 & 28          & 1.00E-04 & 8            & 0.995    & 46.2 & 77.8 & 7.6 & 61.9   \\
 & 28          & 1.00E-04 & 8            & 0.999    & 46.2 & 77.8 & 7.7 & 61.9   \\
\midrule
\multirow{12}{*}{CrySyS}
 & 27          & 3.00E-04 & 5            & 0.95     & 5.3   & 99.1   & 2.7  & 78.1     \\
& 27          & 3.00E-04 & 5            & 0.99      & 2.9   & 99.6   & 1.7  & 78.1     \\
& 27          & 3.00E-04 & 5            & 0.995     & 2.6   & 99.6   & 1.6  & 78.1     \\
 & 27          & 3.00E-04 & 5            & 0.999    & 0.2  & 99.7   & 1.3  & 78.1     \\
 & 30          & 5.00E-05 & 5            & 0.95     & 9.0   & 98.6   & 4.2  & 79.6     \\
 & 30          & 5.00E-05 & 5            & 0.99     & 8.0   & 98.9   & 3.9  & 79.6     \\
 & 30          & 5.00E-05 & 5            & 0.995    & 7.8   & 98.9   & 3.8  & 79.6     \\
& 30          & 5.00E-05 & 5            & 0.999     & 7.5   & 99.0   & 3.7  & 79.6     \\
 & 30          & 1.00E-04 & 8            & 0.95     & 16.9  & 97.7   & 6.4  & 79.1     \\
 & 30          & 1.00E-04 & 8            & 0.99     & 16.1  & 97.9   & 6.2  & 79.1     \\
 & 30          & 1.00E-04 & 8            & 0.995    & 15.5  & 98.0   & 6.0  & 79.1     \\
& 30          & 1.00E-04 & 8            & 0.999     & 11.7  & 98.5   & 5.0  & 79.1    \\
\bottomrule
\end{tabular}
\end{table*}

A common preprocessing step is the aggregation of messages into fixed-length time windows, either by message counts or time range, introducing a critical hyperparameter: the window size. This parameter directly affects the temporal resolution of the data and the label assignment, as multiple messages are aggregated into a single decision unit.
Fig.~\ref{fig:window-analysis} illustrates the impact of varying window sizes on the number of attack and benign windows, as well as the average number of messages per window, for two of the tested datasets.

In the HCRL-CarHacking dataset, both attack and benign window counts decrease gradually as the window size increases from 0.01\,s to 10\,s. In contrast, in the DAGA dataset, the number of benign windows rapidly drops and reaches zero after a window size of 1\,s.
This behavior can be explained by the temporal distribution of messages. In both datasets, attack messages are densely concentrated within time intervals. As the window size increases, these dense regions increasingly dominate entire windows, resulting in most windows being labeled as attacks. The HCRL-CarHacking dataset contains additional benign traffic following the attack period, which explains the persistence of benign-only windows even at larger window sizes. We manually examined the same window size configurations across the remaining datasets and observed similar trends.

Choosing an appropriate window size is indeed a challenging task, as it may depend on factors beyond detection capability, such as processing time, detection delay, and operational constraints. There is a trade-off between having enough messages to capture meaningful patterns and not waiting too long to detect attacks with minimal delay. A commonly used window size typically ranges from a few milliseconds to approximately 3 seconds. Based on this analysis, we used a window size of 1\,s in our experiments, as it provides a balance between sufficient temporal context and a representative distribution of attack and benign samples, with approximately 1600 messages per windows.

\section{Hyperparameter Grid Search}
\label{appendix:grid_search}

\begin{table*}[]

\centering
\caption{Grid search results for the CANShield-CNN-AE method on the ROAD and CrySyS dataset.}
\label{tab:appendix_canshield_grid_search}

\begin{tabular}{lllllllll}
\toprule

\textbf{Dataset} & \textbf{Loss\_factor} & \textbf{Time\_factor} & \textbf{AUC(\%)} & \textbf{bACC(\%)} & \textbf{F1(\%)} &\textbf{MCC(\%)} & \textbf{Precision(\%)} & \textbf{Recall(\%)} \\
\midrule
\multirow{6}{*}{ROAD}   & 0.99           & 0.99           & 75.3                & 77.8                              & 71.5              & 68.1               & 99.6                     & 55.8                  \\
                        & 0.97           & 0.99           & 74.7                & 77.7                              & 71.2              & 67.5               & 98.7                     & 55.7                  \\
                        & 0.99           & 0.95           & 75.8                & 77.3                              & 70.3              & 65.5               & 95.2                     & 55.8                  \\
                        & 0.97           & 0.95           & 75.4                & 77.0                                & 69.9              & 65.5               & 96.4                     & 54.8                  \\
                        & 0.95           & 0.99           & 79.6                & 77.1                              & 69.9              & 64.8               & 94.2                     & 55.6                  \\
                        & 0.95           & 0.95           & 76.7                & 74.3                              & 64.9              & 54.5               & 78.7                     & 55.2 \\
\midrule
\multirow{6}{*}{CrySyS} & 0.99           & 0.95           & 65.0                  & 65.3                              & 30.4              & 20.3               & 18.8                     & 80.5                  \\
                        & 0.99           & 0.99           & 59.4                & 61.1                              & 27.6              & 14.9               & 16.8                     & 75.9                  \\
                        & 0.97           & 0.95           & 57.0                  & 60.1                              & 26.8              & 13.6               & 16.2                     & 77.7                  \\
                        & 0.95           & 0.99           & 59.6                & 59.1                              & 26.1              & 12.7               & 15.6                     & 81.6                  \\
                        & 0.95           & 0.95           & 57.5                & 56.6                              & 24.8              & 10.9               & 14.3                     & 91.4                  \\
                        & 0.97           & 0.99           & 54.4                & 55.6                              & 24.2              & 7.8                & 14.4                     & 74.1                  \\

\bottomrule
\end{tabular}
\end{table*}

In the case of the CANet-LSTM-AE method, we conducted a grid search over key hyperparameters on the ROAD and CrySyS dataset, including the learning rate, hidden size, and decision threshold (Quantile). Table~\ref{tab:appendix_canet_grid_search} summarizes the evaluated configurations and their performance.

For the CrySyS dataset, true positive rates (TPR) remain below 20\% across all configurations, indicating that the model fails to reliably detect attack instances. It is worth noting that a trace-level examination revealed that certain attack scenarios can be detected more effectively, with TPR even reaching up to 75\% and true negative rates (TNR) around 88\%. However, this behavior was not consistent across all attack types.
Results on the ROAD dataset show comparatively higher TPR, however, the overall performance remains limited. Although some attacks can be detected with 100\% TPR and 74\% TNR, the same configuration fails to detect other attack traces, suggesting that detection is still not sufficiently reliable.

Notably, the Area Under the Curve (AUC) remained similar across different setups, indicating that the ranking quality of anomaly scores is largely unaffected by the decision threshold. Increasing the number of training epochs did not improve performance, suggesting that the model has already reached its capacity under the current training setup and does not benefit from further optimization within this configuration. Overall detection performance, as measured by the F1-score, remained low across all configurations, which may be attributed to the more challenging nature of these dataset, as opposed to the SynCAN dataset.

The limited performance gains achieved through hyperparameter tuning suggest that the observed performance gap is unlikely to be solely attributable to suboptimal parameter choices. Instead, this indicates that effectively applying the model to more heterogeneous, real-world CAN data may require additional considerations beyond standard parameter tuning. In particular, adapting the method to such settings appears to be non-trivial and may require more substantial modifications.

Similarly, we tuned CANShield-CNN-AE using a structured grid search performed separately for the ROAD and CrySyS dataset. Table~\ref{tab:appendix_canshield_grid_search} summarizes the evaluated configurations and their performance. The search was designed to study both the temporal granularity of the CANShield-CNN-AE input representation and the strictness of its thresholding procedure.
During preprocessing, a sampling period parameter controls temporal subsampling when constructing each input window. A value of 1 uses every timestep, whereas larger values such as 5 or 10 keep every 5th or 10th timestep, respectively. Consequently, the input tensor always contains 50 sampled points, but larger sampling periods increase the raw temporal span covered by each window. Although the sampling period of 1, 5 and 10 was tested, results did not improve with larger temporal span, and thus in all reported runs it was fixed to 1. 

Consequently, the reported results focus on the hierarchical threshold selection. CANShield-CNN-AE computes signal-wise reconstruction errors for each window. The \textit{Loss\_factor} parameter defines the percentile used to threshold the reconstruction error of each signal individually. After that, the binary exceedances are aggregated over time within a window. The parameter \textit{Time\_factor} defines the percentile threshold applied to the fraction of timesteps in a window for which a given signal exceeds its loss threshold. Thus, \textit{Time\_factor} controls how persistent the abnormal behavior must be over time before a signal is considered anomalous at the window level. CANShield-CNN-AE also uses a third thresholding stage, where anomalies are aggregated across signals, however, this parameter was held fixed at 0.95 in the present experiments and was therefore not included in the reported grid.

For each trial, results report the threshold-free AUC together with balanced accuracy (bACC), F1 score, Matthews correlation coefficient (MCC), Precision, and Recall. 
\section{Datasets selection and comparison}
\label{appendix:datasets}

We selected the evaluation datasets based on two main considerations. First, for the reproduced methods, we included the datasets originally used in their respective papers. Second, based on survey papers, we included the datasets most commonly used in the field.
Although a few other publicly available CAN traffic datasets exist, we selected these datasets because of their popularity and the availability of published methodologies.

Due to differences in dataset characteristics, not all methods could be meaningfully evaluated on all datasets. Since CANet-LSTM-AE and CANShield-CNN-AE target signal-level attacks, we evaluated them separately and only on SynCAN, ROAD, and CrySyS, which contain message modification attacks. In contrast, AssocRules, FlowNGram, and MBA-OCSVM were evaluated not only on OTIDS, CIDv2, CarHacking, and DAGA, but also on CrySyS and ROAD, because these datasets contain message injection attacks in addition to message modification attacks. SynCAN, however, was unsuitable for these three methods because it contains no message injections and does not provide raw CAN traffic.

Thus, the final setup was as follows: AssocRules, FlowNGram and MBA-OCSVM were tested on message injection attacks in the OTIDS, CIDv2, CarHacking, CrySyS, DAGA, and ROAD dataset, while CANet-LSTM-AE and CANShield-CNN-AE were tested on message modification attacks in the SynCAN, ROAD, and CrySyS datasets.

\end{document}